Title :DERIVATION OF THE TULLY-FISHER's LAW FROM THE THEORY OF ETHER
Author:Thierry DELORT
Date 29th August 2011
Email :tdelort@yahoo.fr

Abstract:
The astrophysicist S. Mc Gaugh recently published [3] a result raising questions concerning the existence of dark matter and the validity of General Relativity, conclusion of his observations of 47 galaxies. He established the validity of Tully-Fisher's law for all those galaxies, which is unexplained and in contradiction with classical Cosmology based on General Relativity. Moreover, we remind that this classical Cosmology did not solve the enigma of the nature of dark matter, nor the enigma of its invisibility, nor the enigma of the constancy of the curve of velocities of stars in a galaxy. Consequently it appears that a new Cosmology, alternative to the classical Cosmology, is necessary, at least in the field of dark matter. In fact it exists already a new Cosmology, alternative to classical Cosmology and compatible with the Relativity Theory, which is the Cosmology of the Theory of Ether relative to dark matter. This new Cosmology gives the nature of dark matter, the origin of its invisibility and permits to obtain the constancy of the curve of velocities of stars in a galaxy. In this article, after having recalled the theoretical elements of this new Cosmology, we will show that it also permits to obtain theoretically Tully-Fisher's law whose Mc Gaugh observations confirmed the validity.
We will note that all the theoretical results used and obtained in this article are compatible with the Special and General Relativity Principles.

Key words: Tully-Fisher's law, dark matter.

1.INTRODUCTION

The theory exposed in this article is based on a new Cosmology that is the Cosmology of the Theory of Ether relative to dark matter. Despite of the fact that this new Cosmology has been elaborated inside the frame of a very general Theory of Ether, we will note that it is compatible with the Special and General Relativity Principles.
So in this article, we suppose the validity of the modern Theory of Ether exposed in the book [4] (That contains rectified and reactualized version of the articles [1][5][6][7][8][9][10] that need to be improved). But we will use only theoretical elements compatible with the Relativity Theory that we will recall at the beginning of the article.
In 3 articles [1] [2] [9], we exposed how this new Cosmology based on a Theory of Ether interprets the nature of dark matter. In this interpretation, dark matter is a substance, called *Ether-substance*, which constitutes what is usually called "The vacuum".
Besides, an astrophysicist, Stacy Mc Gaugh, established experimentally in the article [3] the validity of Tully-Fisher's law, showing that the baryonic mass of a galaxy is proportional to the 4th power of the velocity of the stars in this galaxy. This fact appears to be in contradiction with the existence of dark matter, because it implies that the velocity of a star depends only on the baryonic mass of the galaxy, and consequently with General Relativity that needs dark matter to explain that the stars are retained inside galaxies.
In reality, in the classical present Cosmology, we have 2 first enigmas: The nature of dark matter and the origin of its invisibility. The interpretation by the new Cosmology of the nature of dark matter, which is according to it a substance constituting what is usually called "The vacuum", explaining its invisibility, permits consequently to solve the 2 preceding enigmas and is consequently extremely interesting. But we obtain another very amazing result: Modeling the Ether-substance as an ideal gas, we obtained in each article [1] [2] [9] the constancy of the curve of stars in galaxies. The constancy of this curve is also unexplained and constitutes a 3$^{rd}$ enigma in the present Classical Cosmology.

We are going to see in this article how our model of the Ether-substance as an ideal gas permits to solve a 4$^{th}$ fundamental enigma that is the validity of Tully-Fisher's law.

It is clear that the fact that what we call "the vacuum" be constituted of a substance (called Ether-substance) owning a mass is in agreement with Special and General Relativity Principles. We will see that it is also the case for the whole of the new Cosmology used in this article in order to solve the 4 fundamental enigmas concerning dark matter.

2. RECALL

We remind that the Tully-Fisher's law is the following:
Tully and Fisher realized some observations on spiral galaxies. They obtain that the luminosity L of a spiral galaxy is proportional to the 4$^{th}$ power of the velocity v of stars in this galaxy. So we have the Tully-Fisher's law for spiral galaxies, $K_1$ being a constant:

$$L = K_1 v^4 \qquad (1)$$

But the baryonic mass M of a spiral galaxy is proportional to its luminosity. So we have also the law for a spiral galaxy, $K_2$ being a constant:

$$M = K_2 v^4 \qquad (2)$$

This 2$^{nd}$ form of Tully-Fisher's law is known as the *baryonic Tully-Fisher's law*.

We remind that the Tully-Fisher's law (1) is not verified in general for galaxies that are not spiral galaxies. But the observations of Mc Gaugh [3] show that the baryonic Tully-Fisher's law (2) seems to be true for all galaxies. This constitutes a new major enigma for the classical Cosmology, but we are going to see how we can derivate this law from the new Cosmology.

Let us recall how in the Chapter 2.2 Black mass of each article [1] and [2] we solved the enigma of the constancy of the curves of velocities of stars a galaxy:
In order to obtain this result, we first modeled the Ether-substance as an ideal gas: An element of Ether-substance with a mass m, a volume V, a pressure P and a temperature T verifies the law, $k_0$ being a constant:

$$PV = k_0 m T \qquad (3)$$

Which means, setting $k_1 = k_0 T$:

$$PV = k_1 m \qquad (4a)$$

Or equivalently, ρ being the density of the element:

$$P = k_1 \rho \qquad (4b)$$

We then emitted the hypothesis that a galaxy could be modeled as a concentration of Ether-substance presenting a spherical symmetry, at a constant and homogeneous temperature T.

We then considered the sphere S(r) (resp.the sphere S(r+dr)) that is the sphere inside the concentration of Ether-substance with a radius r (resp. r+dr) and whose the center is the center O of the galaxy.
We then considered the following equation (5a) of equilibrium of forces on an element of Ether-substance with a surface dS, a width dr, situated between the 2 spheres S(r) and S(r+dr):

$$dSP(r+dr) + \frac{G}{r^2}(\rho(r)dSdr)(\int_0^r \rho(x)4\pi x^2 dx) - dSP(r) = 0 \qquad (5a)$$

We then verified that the density of the ether-substance ρ(r) satisfying the preceding equation of equilibrium was:

$$\rho(r) = \frac{k_2}{4\pi r^2} \qquad (5b)$$

The constant $k_2$ being given by, G being the Universal attraction gravitational constant:

$$k_2 = \frac{2k_1}{G} = \frac{2k_0 T}{G} \qquad (6)$$

Using the preceding equation (5b), we obtain that the mass M(r) of the sphere S(r) constituted of Ether-substance is given by the equation:

$$M(r) = \int_0^r 4\pi x^2 \rho(x) dx = k_2 r \qquad (7)$$

We then obtain, neglecting the mass of stars in the galaxy, that the velocity v(r) of a star of a galaxy situated at a distance r from the center O of the galaxy is given by $v(r)^2/r = GM(r)/r^2$ and consequently :

$$v(r)^2 = Gk_2 = 2k_1 = 2k_0 T \qquad (8)$$

So we obtain in the previous equation (8) that the velocity of a star in a galaxy is independent of its distance to the center O of the galaxy, solving the 3rd enigma concerning dark matter. We justify in details the equations (5b) and (6) in the Chapter 2.2 Black mass of any of the 2 articles [1] and [2], it is easy to verify them using the equation (4b).
We note that the theoretical elements of the new Cosmology permitting to obtain the equations (5b)(6)(7)(8) are compatible with Special and General Relativity Principles.

3.THEORY OF QUANTIFIED LOSS OF CALORIFIC ENERGY (BY BARYONS)

We saw in the previous equation (8) that according to the new Cosmology, the square of the velocity of stars in a galaxy is proportional to the temperature of the concentration of Ether-substance constituting this galaxy. So if we determine this temperature T, we then obtain the squared velocity of the stars in this galaxy. So we need to try to determine T:
-A first possible idea is that the temperature T is the so called "Temperature of the fossil radiation". But this is impossible because it would imply that all stars of all galaxies are driven with the same velocity and we know that it is not the case.

-A second possible idea is that the temperature T is due to the absorption by the concentration of Ether-substance constituting the galaxy of a fraction of the photons emitted by the stars of this galaxy. But if it was the case, the temperature and consequently the velocity of the stars of the galaxy would only depend on the luminosity of the galaxy, and we should have a law analogous to the Law of Tully-Fisher (1) and we know that it is not the case.
-A third possible idea is that in any galaxy, each baryon interacts with the Ether-substance constituting the galaxy, and consequently it occurs for each baryon a loss of calorific energy communicated to the Ether-substance.
A priori we could expect that this loss of calorific energy for each baryon (transmitted to the Ether-substance) depend on the temperature of this baryon, but if it was the case, the total calorific loss for all baryons would be extremely difficult to calculate and moreover we would not obtain that the total calorific loss depend on the baryonic mass of the galaxy.

The final idea is that indeed it occurs a calorific loss for each baryon (transmitted to the Ether-substance), but that this loss is quantified, depending only on the number of the nucleons of the baryon. This loss should be very low, but the calorific capacity of the Ether-substance being also very low, it can involve an appreciable temperature of the concentration of Ether- substance constituting the galaxy.
So we make the following hypothesis:

HYPOTHESIS OF QUANTIFIED CALORIFIC LOSS (OF BARYONS)

-Each baryon of a galaxy is submitted to a loss of calorific energy, transmitted to the concentration of Ether-substance constituting the galaxy.
-This loss of calorific energy depends only on the number of nucleons constituting the baryon (It is independent of its temperature). So if p is the power corresponding to the loss of calorific energy for a baryon with n nucleons, it exists a constant $p_0$ (loss of calorific energy per nucleon) such that:

$$p = n p_0 \qquad (9)$$

According to the equation (9), the total power corresponding to the loss of calorific energy by all the baryons of a galaxy is proportional to the number of nucleons of the whole of those baryons, and consequently to the baryonic mass of this galaxy. So if $m_0$ is the mass of one nucleon, M being the baryonic mass of the galaxy, we obtain according to the equation (9) that the total power $P_r$ corresponding to the calorific energy received by the concentration of Ether-substance constituting the galaxy from all the baryons is given by the following equation, $K_3$ being the constant $p_0/m_0$:

$$P_r = (M/m_0) p_0 = K_3 M \qquad (10)$$

Concerning the preceding Hypothesis of loss of calorific energy, it is important to remark:
-The loss of calorific energy of a baryon transmitted to the Ether-substance is a quantum phenomenon, consequently it is not surprising that the power corresponding to the loss of calorific energy of a baryon be quantified.
-In physics of thermal transfer, the calorific loss of one or several other particles usually depend on their temperature. But it is always only thermal transfers from baryons to other baryons that are considered, and consequently it is not compulsory that it be also the case for transfers between baryons and Ether-substance.

-It is possible that this hypothesis be true only for baryons whose temperature be superior to a given temperature $T_S$. Moreover, their temperature must be superior to the local temperature of the Ether-substance.
-The great simplicity of this hypothesis permits to obtain very easily the total power corresponding to calorific energy received by the concentration of Ether-substance (Equation (10)). If the loss of energy of a baryon depended on its temperature, then it would be incomparably more complicated, and maybe impossible, to obtain a simple expression giving this total power.
-This hypothesis is a priori compatible with the Special and General Relativity Principles, and also with classical Quantum Physics.

4. DEDUCTION OF THE TULLY-FISHER's LAW

In agreement with the previous model of galaxy, we model a galaxy as a concentration of Ether-substance presenting a spherical symmetry (and consequently being itself a sphere), at a temperature T and immerged inside a medium constituted of Ether-substance at a temperature $T_0$ ($T_0$ being the so called "Temperature of the fossil radiation"), and with a density $\rho_0$.

In order to obtain the radius R of the concentration of Ether-substance constituting the galaxy, it is logical to make the hypothesis of the continuity of $\rho(r)$: R is the radius for which the density $\rho(r)$ of the concentration of Ether-substance is equal to $\rho_0$. So we have the equation:

$$\rho(R) = \rho_0 \qquad (11)$$

Consequently we have according to the equations (5b) and (6):

$$\frac{k_2}{4\pi R^2} = \rho_0 \qquad (12)$$

$$\frac{2k_0 T}{G} \times \frac{1}{4\pi R^2} = \rho_0 \qquad (13)$$

So we obtain that the radius R of the concentration of Ether-substance constituting the galaxy is given approximately by the equation:

$$R = \left(\frac{2k_0 T}{4\pi G \rho_0}\right)^{1/2} = K_4 T^{1/2} \qquad (14)$$

The constant $K_4$ being given by:

$$K_4 = \left(\frac{2k_0}{4\pi G \rho_0}\right)^{1/2} \qquad (15)$$

We can then consider that the sphere with a radius R of Ether-substance constituting the galaxy is in thermal interaction with the medium at a temperature $T_0$ in which it is immerged. We model this thermal interaction as a convection phenomenon. If $\varphi$ is the thermal flow of energy on the borders of the sphere, the power $P_l$ lost by the sphere of Ether-substance constituting the galaxy is given by the equation:

$$P_l=4\pi R^2\varphi \qquad (16)$$

But we know that for a convection phenomenon between a medium at a temperature T and a medium at a temperature $T_0$ the flow $\varphi$ between the 2 media is classically given by the expression, h being a constant depending only on $\rho_0$:

$$\varphi=h(T-T_0) \qquad (17)$$

Consequently the total power lost by the concentration of Eher-substance is:

$$P_l=4\pi R^2 h(T-T_0) \qquad (18)$$

We can consider that at the equilibrium, the thermal power $P_r$ received by the concentration of Ether-substance constituting the galaxy is equal to the thermal power $P_l$ lost by this concentration. Consequently according to the equations (10) and (18), M being the baryonic mass of the galaxy, we have:

$$K_3 M=4\pi R^2 h(T-T_0) \qquad (19)$$

Using then the equation (14) :

$$K_3 M=4\pi K_4^2 hT(T-T_0) \qquad (20)$$

Making the approximation $T_0<<T$ (We remind that $T_0$ is of the order 3°K) :

$$M = 4\pi \frac{K_4^2}{K_3} hT^2 \qquad (21)$$

Consequently we obtain the expression of T, defining the constant $K_5$ :

$$T = (\frac{K_3}{4\pi K_4^2 h})^{1/2} M^{1/2} = K_5 M^{1/2} \qquad (22)$$

And then according to the equation (8) :

$$v^2=2k_0 T=2k_0 K_5 M^{1/2} \qquad (23)$$

So :

$$M = (\frac{1}{2k_0 K_5})^2 v^4 \qquad (24)$$

So we finally obtain :

$$M=K_6 v^4 \qquad (25)$$

The constant $K_6$ being defined by:

$$K_6 = \left(\frac{1}{2k_0 K_5}\right)^2 = \frac{4\pi K_4^2 h}{4k_0^2 K_3}$$

$$K_6 = \frac{4\pi h}{4k_0^2 K_3} \times \frac{2k_0}{4\pi G \rho_0}$$

$$K_6 = \frac{m_0 h}{2k_0 G \rho_0 p_0} \qquad (26)$$

So we obtain the baryonic Tully-Fisher's law (2), with $K_2 = K_6$.

4. CONCLUSION

So we solved the 4th enigma concerning dark matter, meaning that we justified theoretically the validity of the Tully-Fisher's law, using a new Cosmology that is the Cosmology of the Theory of Ether relative to dark matter. We note that the theoretical elements of this new Cosmology in order to solve this 4th enigma are, as all the theoretical elements used in order to solve the 3 other enigmas, compatible with the Special and General Relativity Principles. In order to obtain the baryonic Tully-Fisher's law, we deepened our model of ideal gas for the Ether-substance, we introduced the Hypothesis of quantified loss of calorific energy (for baryons), and we used a simple thermal model for a galaxy. So we see that the presented here new Cosmology, being compatible with Special and Relativity Principles, appears to be much more powerful than the classical Cosmology in the field of dark matter. We remind that despite of its compatibility with the Relativity Theory, it has been elaborated in the frame of a very general Theory of Ether [4][5][6][7][8][9][10].